\documentclass[letterpaper]{article}

\usepackage{natbib,alifeconf}  
\usepackage{csquotes}

\newcommand{\figref}[1]  {Fig.~\ref{#1}}

\title{The modularity of action and perception revisited using control theory\\ and active inference}
\author{Manuel Baltieri$^{1,2}$ \and Christopher L. Buckley$^{1,2}$ \\
\mbox{}\\
$^1$ Evolutionary and Adaptive Systems Group, Department of Informatics, \\ University of Sussex, Brighton, UK \\
$^2$ Sussex Neuroscience, University of Sussex, Brighton, UK \\
m.baltieri@sussex.ac.uk} 

\begin{document}
\maketitle

\begin{abstract}
The assumption that action and perception can be investigated independently is entrenched in theories, models and experimental approaches across the brain and mind sciences. In cognitive science, this has been a central point of contention between computationalist and 4Es (enactive, embodied, extended and embedded) theories of cognition, with the former embracing the ``classical sandwich", modular, architecture of the mind and the latter actively denying this separation can be made. In this work we suggest that the modular independence of action and perception strongly resonates with the \emph{separation principle} of control theory and furthermore that this principle provides formal criteria within which to evaluate the implications of the modularity of action and perception. We will also see that real-time feedback with the environment, often considered necessary for the definition of 4Es ideas, is not however a sufficient condition to avoid the ``classical sandwich". Finally, we argue that an emerging framework in the cognitive and brain sciences, active inference, extends ideas derived from control theory to the study of biological systems while disposing of the separation principle, describing non-modular models of behaviour strongly aligned with 4Es theories of cognition.

\end{abstract}

\section{Introduction}
\begin{displayquote}
Can perception and action be studied as separate processes?
\end{displayquote}
In cognitive science, the hypothesis that the mind is modular originated with Fodor's work \citep{fodor1983modularity}, formalising the idea that the perceptual and motor systems should be considered as separate and informationally encapsulated components of an organism that sit at its periphery. This view has then more recently been recapitulated by the so-called ``classical sandwich'' architecture of cognitive systems, whereby cognition sits in between perception and action, effectively rendering them almost autonomous \citep{hurley2001perception}. This view contrasts with 4Es (enactive, embodied, embedded and extended) theories of the mind suggesting that real-time feedback interactions with the environment are crucial to explain cognitive processes. In doing so, 4Es proposals reject the hypothesis of segregated perceptual and motor components \citep{clark1998being, wilson2002six, beer2015information, di2017sensorimotor}, now seen as strongly and reciprocally coupled by feedback mechanisms mediated by the environment.

In this work we argue, however, that the presence of feedback is not enough to reject the classical sandwich architecture. We will see that the modular view can still implicitly survive in modern studies of action and perception even in the presence of closed sensorimotor loops. To ground this argument and related discussions on Fodor's modularity, we use formal frameworks that have emerged from control theory, already widely exploited in modern theories of perception and action based on processes of estimation/inference and control \citep{rao1999predictive, todorov2004optimality, friston2010free, friston2011optimal}. Perception, on this view, is modelled as a process of estimation of the hidden or latent states of the world given noisy and often inaccurate observations. Action, on the other hand, is accounted for with theories of optimal control, suggesting possible optimality principles for the implementation of motor actions and behaviour more in general. In this light, we then argue that the so-called \emph{separation principle} of estimation and control in control theory provides concrete grounding of Fodor's modularity with regards to action and perception. After presenting this principle and its connections to the idea of modularity, we will discuss its weaknesses for the study of cognitive and natural systems. We finally propose active inference \citep{Friston2010biocyb, friston2010free} as an alternative view openly rejecting the separation principle, thus supporting non-modular 4Es arguments, while maintaining its roots in modern approaches to action and perception based on control theory.

\section{Perception and action in cognitive science}
\subsection{The classical sandwich}
A classical idealisation of perception holds that it is a bottom-up or feed-forward process with the primary goal of receiving information through the senses in order to build internal representations of the surrounding environment \citep{marr1982vision}. Cognition, including thinking, planning, etc., is cast as a process of manipulating the information within these internal representations and action as a process of deriving appropriate motor commands based on the outcomes of those manipulations. This sequential architecture of perception-cognition-action is best captured by the idea of the ``classical sandwich'' \citep{hurley2001perception}. On this view, perceptual and motor processes work relatively independently, separated by cognitive manipulations at their centre. This notion of separation has its roots in Fodor's idea of the modular mind \citep{fodor1983modularity}. Fodor saw the more traditional notion of cognition, the \emph{filling}, representing functions such as memory, problem solving, deduction and induction, as strictly non-modular. However, his argument for \emph{peripheral} processing, input interfaces (i.e. perception) and output layers (i.e. motor control, action), is that these components should be seen as separate modules. Modules have only restricted access to higher order information and vice versa, and their information content is encapsulated, limited and specific to each module. According to this view then, expectations, beliefs and desires also cannot affect in particular perceptual systems \citep{fodor1983modularity, barrett2006modularity} (although we note that the exact meaning of Fodor's idea is still debated, \cite{coltheart1999modularity, barrett2006modularity}).

\subsection{The 4Es view}
In contrast, enactive, embodied, extended and embedded theories of cognitive science, often just addressed as 4Es, challenge several of the intuitions behind computational accounts of cognitive science, including the sequential and modular nature of perception, cognition and action proposed with the classical sandwich \citep{varela1991embodied, clark1998being, wilson2002six, beer2015information}. 4Es theories cover a large set of heterogeneous ideas with some of the more general points including the importance of fast-paced, dynamic interactions with the environment over internal computations and the role played by the body in this dynamical process. The dynamic nature of real-time interactions of an agent with its environment leads then to a non-sequential, non-encapsulated interpretation of perception and action, best represented as a causally circular (i.e. closed) process whose components are not clearly separable. This last idea is the point of contention we focus on in this work. We claim, in fact, that even in closed-loop frameworks including feedback mechanisms, the often implicit assumption is that perception and action can still be treated separately (for counterexamples see for instance \cite{beer2003dynamics, iizuka2004simulating, harvey2005evolutionary, di2008not}). This becomes especially clear in the literature focused on notions of optimal control, whereby definitions of optimality often include assumptions regarding the separation of perceptual and motor components within a system. In the next section we will see how concepts from control theory, estimation, optimal control and in particular the \emph{separation principle}, can explain this apparent paradox (i.e. closed-loop but nonetheless optimally separable).

\section{Perception and action in control theory}
\subsection{Perception as inference (estimation)}
In control theory, the idea of estimation first emerged with the contributions of Kolmogorov and Wiener \citep{sorenson1970least} and the popularisation of Kalman filters \citep{kalman1960new}, although estimation theory itself can be traced back to Gauss and his method of least squares \citep{sorenson1970least}. The main goal of estimation is to infer (or estimate) hidden parameters, states and inputs of system given only a set of observations that are in general noisy, and a model of the underlying dynamics that may also include noise/uncertainty.

More recently, these ideas have re-emerged in the context of studies of \emph{perception as inference} with the so-called ``Bayesian brain hypothesis'' and related theories of predictive coding and free energy minimisation \citep{knill1996perception, rao1999predictive, knill2004bayesian, friston2010free, hohwy2013predictive, clark2015surfing, buckley2017free}. On this view, organisms operate in an uncertain world, with noisy sensors that provide only incomplete and often ambiguous information. Agents are depicted as Bayesian inference machines that estimate the latent states and causes of their sensory input via the update of a generative model of such input. These agents operate by generating top-down predictions of their sensory data and by continuously updating these prediction to minimise their difference to incoming sensory data (prediction errors). Once prediction errors are minimised, they converge to the best explanation, or inference, of the causes of their sensory data. In control theory, this process of inference is often defined as ``estimation'' or ``filtering'' \citep{jazwinski1970stochastic, astrom2010feedback, astrom2012introduction} and is usually performed by an ``estimator''. In the literature, an estimator is sometimes considered to include a forward model, or predictor \citep{todorov2006optimal}, in other instances to work alongside one (see discussion in \cite{friston2011optimal}). In our work we will consider an estimator-predictor pair as simply an estimator.

\subsection{Action as control}
Modern optimal control theory originated in 1950-60's with formalisations by Pontryagin \citep{pontryagin1962mathematical} and Bellman \citep{bellman1957dynamic}. Connections to classical mechanics and to work by Hamilton and Jacobi among others \citep{todorov2006optimal, sussmann1997300} however place its historical inception a few centuries earlier. Optimal control defines the problem of finding a policy (i.e. a sequence of actions) for a given system that optimises a criterion describing a certain goal for the system.

In the last few decades, optimal control has also emerged as a dominant theory of action, motor control and behaviour in neuroscience \citep{kawato1999internal, wolpert2000computational, todorov2004optimality, kording2006bayesian, kording2007decision, franklin2011computational}. On this view, organisms minimise, over time, a cost function (i.e. the optimality criterion) representing a measure of performance for the achievement of a certain goal. For instance, smooth reaching hand movements can be explained by the minimisation of a cost function based on the rate of change of acceleration, or jerk, of hand movements \citep{franklin2011computational}.

Different cost functions lead to different possible control policies (i.e. sequences of actions) towards a goal. One of the most fundamental distinctions between classes of cost functions is based on the absence/presence of real-time feedback from the environment, defining open and closed-loop control respectively \citep{todorov2004optimality, astrom2010feedback}. Open-loop control methods rely on complex internal models that accurately mimic the external dynamics simulating also the effects of feedback, and allow for pure internal feed-forward planning. Perceptual processes model the transduction of sensory input into some internal neural representation and are often represented as estimators \citep{todorov2004optimality} and/or forward models-estimators pairs \citep{wolpert1998multiple}. On the other hand, action corresponds to the motor output produced by such representations and is usually portrayed as a controller \citep{todorov2004optimality} or inverse model \citep{wolpert1998multiple, kawato1999internal}. The sequential nature of estimation-modelling (planning)-control is consistent with the traditional classical sandwich of cognitive science, where estimation and control are processes encapsulated into separate modules used in sequential order and the effects of feedback from the environment are though to be slow enough to be successfully modelled internally. Closed-loop control is based on the same architecture as the open-loop case (i.e. estimators and controllers) but includes, in contrast, fast-paced feedback from the environment. This allows such framework to tackle more elegantly different sources of noise, delays, internal fluctuations and uncertainties \citep{todorov2004optimality}. Closed-loop control appears to be closer to 4E views with the presence of a real-time feedback mechanism that highlights the dynamic interactions of a system with its environment and the impossibility to capture it internally. We however argue that the most common implementations of closed-loop optimal control are still more consistent with the more traditional, sequential, view of action and perception mainly due to ideas based on the \emph{separation principle}. The vast majority of examples where this sequential view is not explicitly taken, in fact, often include closed-loop control without a specific notion of optimality \citep{beer2003dynamics, iizuka2004simulating, harvey2005evolutionary, di2008not}.

\subsection{The separation principle}
The \emph{separation principle} of control theory provides a set of necessary and sufficient conditions under which an optimal controller can be constructed by combining independent designs for an optimal estimator and an optimal controller \citep{wonham1968separation, astrom2010feedback, georgiou2013separation}. This methodology is widely adopted in control theory and can be used to build controllers for noisy and uncertain systems where environmental states are only partially observed. Separating estimation and control is practically desirable because it becomes then possible to \emph{optimally} solve the estimation problem and subsequently use the output estimate to build an \emph{optimal} controller. Typical designs utilise Kalman filters (estimators) and LQR (Linear Quadratic Regulator) controllers \citep{astrom2010feedback}. The idea of the separation theorem is also very closely related to the certainty equivalence principle described in econometrics \citep{simon1956dynamic, theil1957note, bar1974dual, astrom2012introduction}.

\begin{figure}[ht!]
  \centering
  \includegraphics[width=1\linewidth]{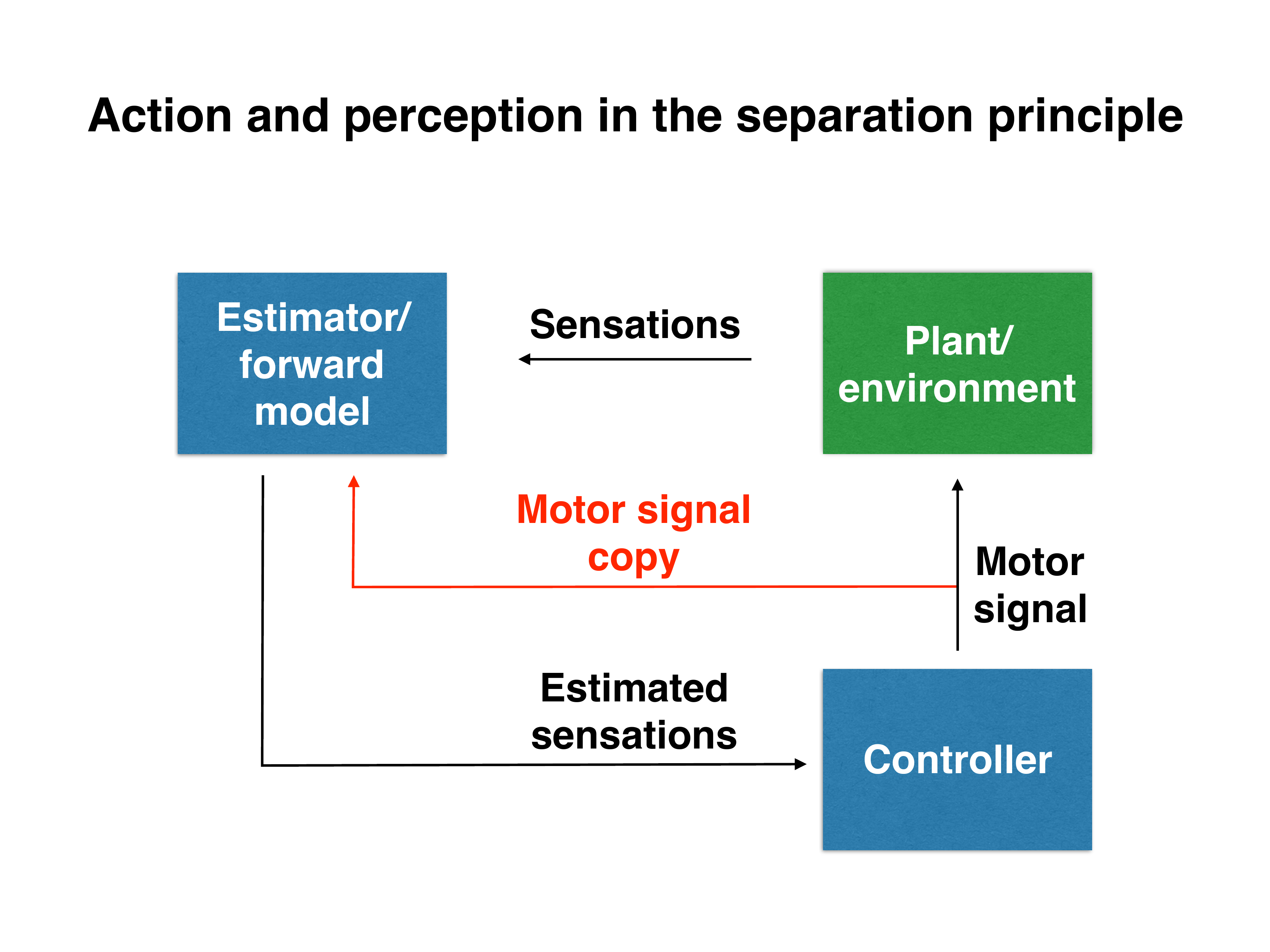}
  \caption{\textbf{A typical control architecture obeying the separation principle.} An estimator, or forward model, infers the causes of incoming sensory input which are relayed to the controller. The controller calculates the optimal output motor signal based on these estimates and allows the system to act on the environment. In parallel, the controller sends also a copy of the command to the estimator, allowing the latter to take this command into account during the estimation of observed stimuli and discount the effects of internally generated actions.}
  \label{fig:EfferenceCopy}
\end{figure}

The separation principle rests on several assumptions which can be summarised by \citep{kappen2011optimal, astrom2012introduction}:
\begin{itemize}
  \item linear process dynamics and observation laws describing the environment and its latent variables
  \item Gaussian noise in both process and measurement equations/laws
  \item known (co)variance matrices representing uncertainty of both process and measurement noises
  \item quadratic cost function used to measure the performance of a system
  \item known inputs for the estimator, meaning that the estimator needs to have access to all the variables, external and internal ones, affecting the inference process.
\end{itemize}
The restrictive nature of these requirements has been widely debated in the control theory literature \citep{astrom2010feedback, astrom2012introduction}. However, here we discuss on their possible meaning for the study of cognition with a focus on biological systems, extending some previous critiques \citep{simpkins2008optimal, ponton2016effects}. It is immediately clear it would be hard to argue for the plausibility of these assumptions for biological systems. Living organisms are highly nonlinear and the environments they operate within are also themselves highly nonlinear, thus they cannot be fully understood with systems of linear equations. There is no direct evidence for fluctuations in physical systems to be accurately described by Gaussian random variables, and whether biological system could effectively keep an updated estimate of the uncertainty in environmental variables (represented by covariance matrices in control theory) is also unclear \citep{astrom2010feedback}. It is then not clear if it is possible to accurately describe important biological phenomena with quadratic cost functions \citep{franklin2011computational}. Lastly, the separation principle suggests that sensory systems must have access to an accurate copy of outgoing motor commands. Specifically, as shown in \figref{fig:EfferenceCopy}, under this scheme the estimator and the controller exchange information in two ways. The estimator relays accurate estimates of world variables to the controller, which in turn sends a copy of the motor command back to the estimator. This copy of the motor command is crucial to allow the sensory system to discount sensory consequences of motor actions. In the absence of this information, estimates of world variables quickly become imprecise and subsequently controls become unstable \citep{friston2011optimal}. This notion of a copy of motor signals is consistent with the classical idea of efference copy in neuroscience \citep{von1950reafferenzprinzip, crapse2008corollary, straka2018new}, whose exact definition and plausibility in neural system have been comprehensively challenged \citep{feldman2009new, friston2011optimal, feldman2016active}.

\begin{figure}[ht!]
  \centering
  \includegraphics[width=1\linewidth]{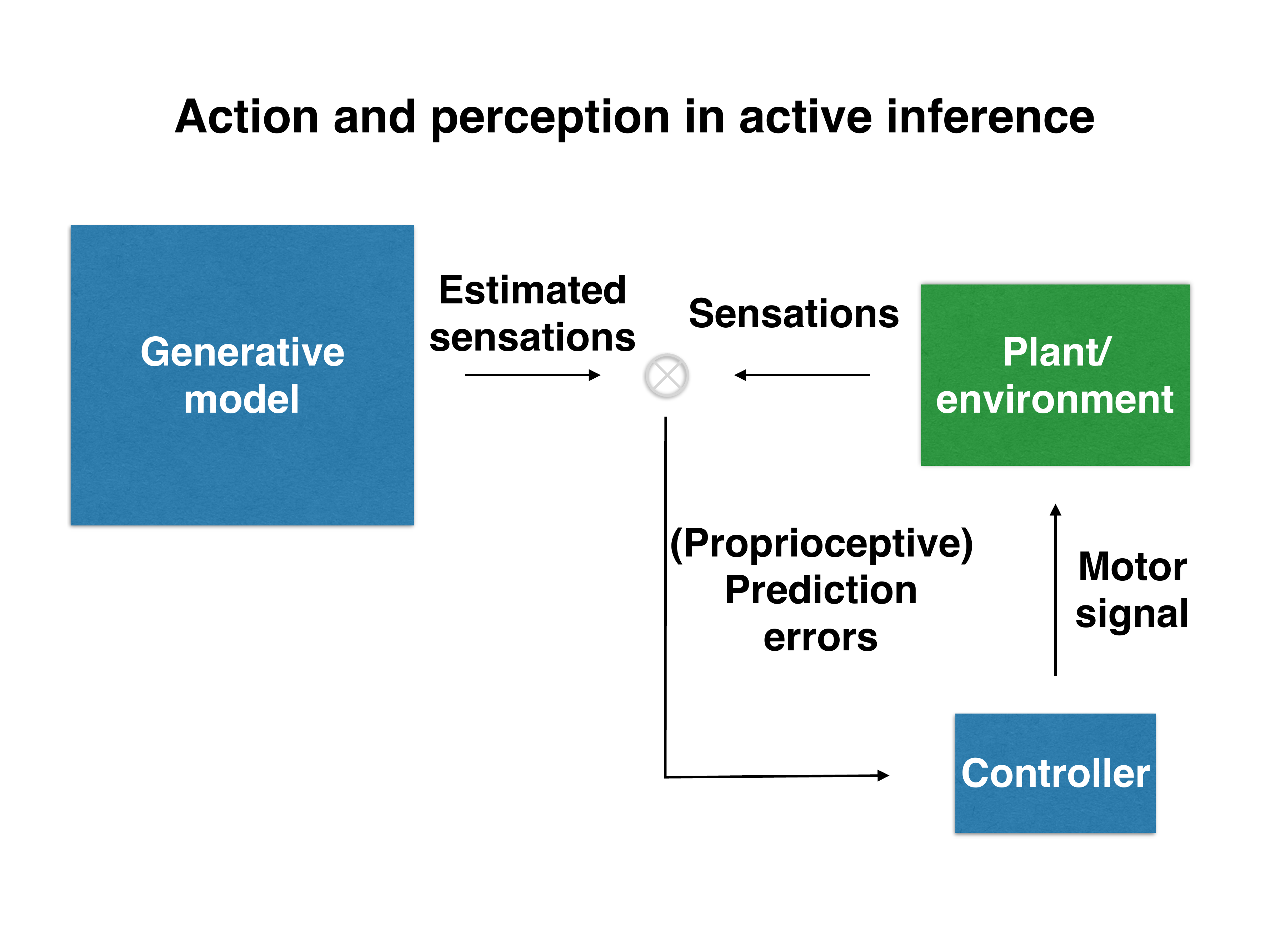}
  \caption{\textbf{A control architecture based on active inference.} Active inference converts the complex problem of optimal control into a more viable problem of inference, solved by a more general generative model \citep{friston2011optimal}. The forward, or generative, model produces estimates of the sensory input. The mismatch between these estimates and real sensory data (here represented as only proprioceptive but more in general including also exteroceptive and interoceptive) generates prediction errors that are used to update the generative model itself and thus infer the causes of sensory data when they are minimised. Propriocetive prediction errors are also explicitly minimised via simple reflex mechanisms implemented at the level of peripheral ``controllers''. These controllers receive information in an intrinsic frame of reference, proprioceptive signals within an agent, and translate them into controls in an extrinsic one, motor actions in the world, using hardwired sensorimotor mappings \citep{Friston2010biocyb}.}
  \label{fig:ActiveInference}
\end{figure}

While the separation principle may not strictly hold, one could argue for a weaker version of separability/modularity. Systems could be cast as ``approximately'' or ``partially'' separable, with such notions still useful metaphors to understand perception and action as separate modules even without optimality as defined by the separation principle. This is especially true for the first four criteria of the separation principle that we listed. For instance, it may be possible to approximate nonlinear descriptions of a system with linearisations around relevant points/equilibria or to estimate covariance matrices that although not optimal, closely resemble the uncertainty of a process. On the other hand, we argue that a notion of ``approximate'' separability is not well defined for the last of the requirements of the separation principle and thus will be the focus of our analysis on the role of this principle in cognitive science. The fact that the estimator needs have information about motor actions of the environment is often not considered a problem in standard control theory and robotics: a copy of motor signals can easily be retrieved and sent back to the estimator/forward model \citep{kawato1999internal}. In biology and neuroscience, however, while the presence of information flowing from motor to sensory areas has been established for decades in the form of efference copy/corollary discharge \citep{cullen2004sensory, crapse2008corollary, straka2018new}, recent discussions on the information contents of such mechanisms \citep{feldman2009new, friston2011optimal, feldman2016active} lead us to carefully consider frameworks based on these ideas and their role in the cognitive and natural sciences. An alternative approach, disposing with the need of a copy of motor signals, is proposed with active inference. In active inference this need is bypassed using a more powerful forward or generative model and trivial sensorimotor mappings in place of complex inverse models/controllers (see \figref{fig:ActiveInference}), as we shall see in the next section.

\section{Perception and action in active inference}
The Free Energy Principle (FEP) is one of the most ambitious attempts to date of unifying several aspects of our understanding of biological systems into a single theory describing such systems in terms of the minimisation of a single quantity: information surprisal \citep{friston2010free, clark2013whatever, hohwy2013predictive, clark2015surfing, buckley2017free}. Surprisal is the negative log-probability of an outcome and for biological systems it is thought to measure the fitness of an organism to specific states, e.g. for a fish, states in the water have low surprisal, and are more valuable than states outside of the water, having high surprisal. Predictive coding and active inference are amongst the most popular approaches derived from the FEP, and constitute process theories proposing how surprisal minimisation could be implemented in living systems. Predictive coding was introduced as a functional model of the visual cortex \citep{rao1999predictive} and then shown to be (computationally) a special case of the FEP \citep{friston2010free}. It effectively implements ideas related to the Bayesian brain hypothesis discussed in previous sections, suggesting that perception is as a process of Bayesian inference and that organisms ``perceive'' the world by minimising prediction errors measuring the difference between internally generated predictions of external stimuli and actually sensed ones. Active inference \citep{Friston2010biocyb, friston2010free} extends predictive coding proposals to the more general realm of action and behaviour. On this view, organisms are not only trying to correct mismatch errors via internal updates of a generative model and its predictions, but also act in the world in order to generate sensations that are better described by existing predictions. Similarly to cybernetics \citep{wiener1961cybernetics, ashby1957introduction}, active inference attempts to connect methods from information and control theory (Variational Bayes and optimal control respectively) to the study of cognition and biology, using empirical evidence to constrain complex theoretical problems \citep{seth2014cybernetic}.

One of the main goals of the FEP is to create a general theory of life and mind, unifying existing knowledge in different fields including thermodynamics, natural sciences, information and control theory \citep{friston2010free, friston2011optimal, friston2012free, friston2013life}. Its position with regards to debates in cognitive science is, however, still unclear. The FEP is often thought to align with more traditional views of cognition, deeply representationalist/computationalist \citep{froese2013brain, gladziejewski2016predictive}. Part of the cognitive science community sees this as a possible advantage \citep{hohwy2013predictive} while others as its main drawback \citep{gladziejewski2016predictive, zahavi2017brain}. Others have argued it is more consistent with a 4Es perspectives of cognition, claiming that the strengths of the FEP reside in generative models with no explicit representational features \citep{bruineberg2016anticipating}. A different perspective highlights the potential of the FEP for the formalisation of ``action-oriented'' views of cognition \citep{engel2016pragmatic, clark2015radical, clark2015surfing, allen2016cognitivism, pezzulo2017model}, re-conciliating traditional views and 4Es theories. In our work we argue that one of the hypotheses expressed by active inference openly rejects one of the requirements of the separation principle that we claim strongly resonates with the ideas of sensory/motor modularity in traditional views of cognitive science. We thus hold that active inference aligns with at least some views from 4Es theories cognition, in particular with the idea of perception and action as deeply entangled functions of embodied and situated agents \citep{clark1998being, wilson2002six, beer2015information, di2017sensorimotor}.

\subsection{Action and perception are not separable}
The active inference framework claims that perfect knowledge of one's motor signals is not necessary in models of control and estimation (and possibly not at all present in biological systems, \cite{friston2011optimal}). The proposed alternative entails recasting motor control problems into perceptual or inference problems, considering that these two classes of problems can be solved by the same algorithms \citep{attias2003planning, todorov2008general, friston2011optimal}. According to this interpretation, perception and action are largely overlapping processes sharing most of their computation, with differences arising mainly at a physiological level. In active inference, the problem of finding actions is essentially converted to an inference problem, solved by the same underlying predictive coding scheme already in charge of perceptual processes. More specifically, in this architecture, proprioceptive sensations are also predicted by an agent, alongside exteroceptive and interoceptive ones (not shown for simplicity in this figure). Explicit motor output is then produced by simple sensorimotor mappings implemented at the very periphery of a system and translating proprioceptive predictions into actions for the external world \citep{Friston2010biocyb}. Conceptually, active inference disposes with the need of a copy of motor signals proposing a more general predictive coding scheme coupled to simple sensorimotor mappings translating proprioceptive predictions into actual actions.

In active inference, the more traditional, sequential and modular role of perception and action advocated by the separation principle is thus questioned, suggesting that these functions are deeply intertwined \citep{friston2011optimal, pickering2014getting, engel2016pragmatic, wiese2016action, pezzulo2017model}. In support of this idea and following our own argument on the parallelism between Fodor's idea of modularity and the separation principle of control theory, we claim that active inference does not meet the requirements for the separation of estimation and control (perception and action). Furthermore, it is engaging in an explicitly non-modular architecture of cognitive processing. We argue, in fact, that while the first four requirements of the separation principle expressed earlier may be prone to arguments regarding separation in at least some approximate sense (e.g. an approximately linear model, noise that is approximately Gaussian, etc.), the presence of a copy of motor signals is either present or not, with no room for approximation. In this sense, the separation of estimation and control is also either present or not. Active inference models explicitly eschew the idea that a copy of motor signals is sent to estimators \citep{friston2011optimal}, an idea strictly necessary for classical motor control architectures based on the separation principle. Without a copy of motor signals, the architecture described by the separation principle is unavoidably broken and thus, according to our initial claim regarding the cognitive sciences, Fodor's modularity cannot be implemented in resulting models of active inference, with action and perception intimately entangled in a non-modular way, only namely ``separated'' for a definition more consistent with traditional views of cognitive science.

\section{Discussion and Conclusions}
In this work we have made an attempt to explicitly connect the idea of modularity in cognitive science to mathematical frameworks of control theory. Specifically, we have proposed that the idea of ``modularity'' \`a la Fodor \citep{fodor1983modularity}, central to architectures such as the classical sandwich of cognitive science \citep{coltheart1999modularity, hurley2001perception, barrett2006modularity}, can be usefully formalised using the separation principle of control theory \citep{wonham1968separation, astrom2010feedback, georgiou2013separation}. In the classical sandwich view, perception and action are seen as modules of a feed-forward-only architecture that are explicitly separated by cognition, the sandwich's ``filling'' \citep{hurley2001perception}. Traditional views of cognitive science openly embrace this architecture and the idea of modularity of perception and action while 4Es (enactive, embodied, embedded, extended) theories largely reject them claiming that fast-paced dynamic interactions between an agent and its environment imply that perception and action are deeply entangled and therefore not modular since such dynamics cannot be internally modelled \citep{varela1991embodied, clark1998being, wilson2002six, di2017sensorimotor}.

To ground the debate arising from these contrasting views, we have then proposed to use control theory, following a general trend in the fields of cognitive science, neuroscience and psychology to adopt theories of estimation and control to explain perception and action respectively \citep{knill1996perception, rao1999predictive, kawato1999internal, todorov2004optimality, franklin2011computational}. All these proposals, however, suggest that these two processes can be treated separately, even when the presence of feedback mechanisms would suggest otherwise. In particular, we claimed that the closed-loop optimal control models largely used in the literature nowadays are often implicitly based on the ``separation principle'' of control theory, with most of the exceptions in work not mentioning explicitly optimality (e.g. \cite{harvey2005evolutionary, di2008not}). This principle defines the conditions whereby a separation of estimation (perception) and control (action) of a system is not only possible, but optimal according to a list of criteria.

We have then claimed that such a separation principle is consistent with the idea of modularity expressed by Fodor, based on the information encapsulation argument \citep{fodor1983modularity, coltheart1999modularity}. Even if the concept of modularity itself is often considered vague \citep{coltheart1999modularity, barrett2006modularity} (and Fodor himself defining modularity only \emph{``to some interesting extent''}, \cite{fodor1983modularity}), we believe that information encapsulation \citep{fodor1983modularity, coltheart1999modularity} should be considered a necessary requirement for modularity. Such encapsulation defines two basic conditions for the existence of a module: (1) restricted access to higher order information and vice versa, and (2) information content limited and specific to each module, both in agreement with the separation principle.

The definition of separation implied by the separation principle in control theory is on the other hand extremely strict and allows for, we suggest, deeper discussions regarding the idea of modularity. More in detail, this principle can be applied only to a small subset of systems (linear, with Gaussian noise, quadratic cost functions, known covariances and known inputs). It is thus hard to imagine, on this view, how studies of brains and minds could make such assumptions, suggesting then that non-modular 4Es views provide a more suitable framework for investigating cognitive and natural systems. We then argued in favour of a recent proposal based on theories of estimation/inference and control and with no explicit assumption regarding their separability: active inference. In this framework, one of the five necessary requirements for separability, the idea of having access to all inputs and in particular to a copy of motor signals for perceptual systems, is dismissed \citep{Friston2010biocyb, friston2011optimal}. By rejecting such mechanism, active inference effectively challenges classical architectures based on the separation principle and in doing so, we claimed, explicitly agrees with 4Es views of cognition whereby perception and action are seen as non-modular processes. Our work provides thus support for hypotheses highlighting how active inference is more in agreement with 4Es theories than with traditional accounts of cognition \citep{clark2015surfing, bruineberg2016anticipating, pezzulo2017model}.

In the future, we will focus on a mathematical treatment of the ideas presented in this manuscript, explicitly highlighting the differences between architectures inspired by the separation principle and proposals more in line with 4Es theories, such as active inference. In doing so, we will also attempt to operationalise our claims regarding biological systems, providing concrete criteria and proposing experimental setups to test our hypotheses (e.g. regarding efference copy/corollary discharge), to disambiguate modularity in living organisms. Another potential contribution will include a deeper analysis regarding aspects of the free energy principle/active inference that still cast doubts on its adherence to 4Es theories. For instance, on the role of the brain as a detached system in a internal/external dichotomy \citep{froese2013brain} or more in general the necessity of markov blankets \citep{friston2013life} for the separation of internal (agent) and external (environment) states that may, almost naturally, introduce the idea of mental representations \citep{allen2016cognitivism}.



Finally, to quote Kalman on an early intuition regarding the problem of simultaneous estimation and control (perception and action in our interpretation):
\begin{displayquote}
\emph{One may separate the problem of physical realization into two stages: computation of the “best approximation” $\hat{x}(t_1)$ of the state from knowledge of $y(t)$ for $t \le t_1$ and computation of $u(t_1)$ given $\hat{x}(t_1)$.}
\end{displayquote}
\begin{flushright}
\citep{kalman1960contributions}
\end{flushright}
This may true for engineering but perhaps not for studies of cognition and natural systems.

\footnotesize
\bibliographystyle{apalike}

\end{document}